\documentclass[prd,showpacs]{revtex4}
\usepackage{amsmath}
\usepackage{epsfig}
\usepackage{amsfonts}
\begin{document}
\def\b{\bar}
\def\d{\partial}
\def\D{\Delta}
\def\cD{{\cal D}}
\def\cK{{\cal K}}
\def\f{\varphi}
\def\g{\gamma}
\def\G{\Gamma}
\def\l{\lambda}
\def\L{\Lambda}
\def\M{{\Cal M}}
\def\m{\mu}
\def\n{\nu}
\def\p{\psi}
\def\q{\b q}
\def\r{\rho}
\def\t{\tau}
\def\x{\phi}
\def\X{\~\xi}
\def\~{\widetilde}
\def\h{\eta}
\def\bZ{\bar Z}
\def\cY{\bar Y}
\def\bY3{\bar Y_{,3}}
\def\Y3{Y_{,3}}
\def\z{\zeta}
\def\Z{{\b\zeta}}
\def\Y{{\bar Y}}
\def\cZ{{\bar Z}}
\def\`{\dot}
\def\be{\begin{equation}}
\def\ee{\end{equation}}
\def\bea{\begin{eqnarray}}
\def\eea{\end{eqnarray}}
\def\half{\frac{1}{2}}
\def\fn{\footnote}
\def\bh{black hole \ }
\def\cL{{\cal L}}
\def\cH{{\cal H}}
\def\cF{{\cal F}}
\def\cP{{\cal P}}
\def\cM{{\cal M}}
\def\ik{ik}
\def\mn{{\mu\nu}}
\def\a{\alpha}

\title{Regularized Kerr-Newman Solution as a Gravitating Soliton}

\author{Alexander Burinskii}

\affiliation{Theoretical Physics Laboratory, NSI, Russian Academy of Sciences,
B. Tulskaya 52  Moscow 115191 Russia\footnote{E-mail address: bur@ibrae.ac.ru}}

\begin{abstract}
\noindent The charged, spinning and gravitating soliton  is
realized as a regular solution of the Kerr-Newman field coupled
with a chiral Higgs model. A regular core of the solution is
formed by a domain wall bubble interpolating between the external
Kerr-Newman solution and a flat superconducting interior. An
internal electromagnetic (em) field is expelled  to the boundary
of the bubble by the Higgs field.  The solution reveals two new
peculiarities: (i) the Higgs field is oscillating, similar to the
known oscillon models, (ii) the em field forms on the edge of the
bubble a Wilson loop, resulting in quantization of the total
angular momentum.
\end{abstract}
\pacs{11.27.+d, 04.20.Jb, 04.70.Bw} \maketitle

{\bf 1. Introduction.}

\bigskip
There are many pieces of evidence that black holes (BHs) may be
related with elementary particles \cite{Sen}. Being regularized,
BHs turn out to be similar to gravitating solitons
\cite{Shnir,Volk} which are considered as regular field models of
the extended particles in the electroweak sector of the standard
model \cite{Dash,Kus,Volk,Mant}.
  The Kerr-Newman (KN) solution plays a fundamental role as a model
  of a
rotating BH and as a convenient laboratory for the study of
consistency between  quantum theory and gravity
\cite{SHW,BurA,BurPreQ}. The aim of this paper is to obtain a
regular version of the KN solution which may also be considered as
a field model of spinning particle compatible with general
relativity.

The attempts to incorporate spin and gravity in the soliton models
encounter the problems related with choosing appropriate field
models. As a result, the models turn out to be difficult for
analysis and admit only numerical solutions
\cite{Shnir,BriHar,SeiSuen}. In this paper we use an alternative
approach starting from a gravitational side --  specific structure
of the KN solution. Our previous attempts to obtain its regular
version \cite{BurBag,BEHM} were based on the bag-like Higgs model
\cite{BurBag} and the G\"urses and G\"ursey (GG) ansatz for the
regular Kerr-Schild (KS) metrics \cite{GG,BEHM}. However, there
were problems of consistency in the electromagnetic (em) sector,
and subsequent attention was drawn to the models based on
non-linear electrodynamics (NED), where the existence of the
regular charged non-rotating BH solutions  was explicitly
demonstrated (in particular in \cite{BurHild0,Dym0}).
Nevertheless, the investigation of the corresponding {\it
rotating} solutions led to contradicting results.\fn{The based on
NED regularized KN solutions were presented in \cite{Dym}, while
our attempts to get such solutions in the NED theory with two
invariants (\cite{BurHild}) indicated their inconsistency with the
GG form of metric (unpublished).} In the present treatment, we
return to the bag-like model considered in \cite{BurBag}  based on
the Higgs mechanism of broken symmetry and show that the problems
in the em sector of the bag-like Higgs model  may be successfully
resolved, leading to new important physical results.

  It was obtained by Carter in 1968 that the KN solution has $g=2$ as that of
 the Dirac electron \cite{DKS}, and there followed a series
 of the works on the KN solution as a model of  a spinning particle, in
 particular \cite{DKS,Isr,Bur0,Lop,BurSen,BurTwi,BurBag,Dym,AP,TN,BurKN}.
  Since the spin of an electron is very high, the BH horizons of the KN solution
disappear, opening a naked singular ring which should be replaced
by a regular source. We consider here a regular field model of the
KN source, in which the KN solution is modified near singular ring
and replaced by a source forming the bubble which has  a
superconducting internal vacuum state formed by the Higgs
mechanism of the broken symmetry. The Higgs sector of our field
model is similar to that of the well-known Nielsen-Olesen (NO)
model for a regular vortex in superconducting media \cite{NO}.
However, to describe a domain wall interpolating between the exact
external KN solution and the flat superconducting interior, we
need the Higgs model containing a few chiral fields. We found that
the regular internal solution is unambiguously determined by the
external KN field, and the regularized KN solution acquires two
important peculiarities:

(i) the internal Higgs field turns out to be spinning and
oscillating with frequency $2m ,$ similar to the models of
oscillating Q-balls \cite{VolkWohn,Grah} and bosonic stars
\cite{BosStar}.

(ii) the external em field acquires on the edge of the bubble a
closed Wilson loop, which entails quantization of angular
momentum.

\bigskip
{\bf 2. The Kerr singular ring, twosheetedness and the bubble
source.}

\bigskip
The problem of regular source of the KN solution becomes
complicated by twosheetedness of the Kerr geometry. The KS form of
metric $ g_\mn=\eta _\mn + 2 H k_\m k_\n $ is coupled to  the
auxiliary Minkowski metric $\eta^\mn $ with the Cartesian
coordinates $x^\m=(t,x,y,z).$ For the KN solution function $H$ has
the form \be H=\frac {mr - e^2/2}{r^2 + a^2 \cos ^2 \theta}
\label{HKN}, \ee  where Kerr's oblate spheroidal coordinates
$r,\theta$ and $\phi_K $ are  related with the Cartesian
coordinates as follows \be x+iy = (r + ia) \exp \{i\phi_K \} \sin
\theta , \ z=r\cos \theta \label{oblate}.\ee

The coordinate $r$ covers the Kerr space-time twice, for $r>0$ and
for $r<0 ,$ forming the `positive' and `negative' sheets which are
analytically connected via disk $r=0.$ On the boundary of the
disk, $r=\theta =0 ,$ function $H$ is singular, which corresponds
to the Kerr singular ring.\fn{Recently, this twosheetedness found
new holographic interpretation \cite{BurA}.} To remove the Kerr
singularity and twosheetedness, they have to be covered by a
material source.  Israel  suggested  to truncate the negative KN
sheet along the disk spanned by the Kerr ring and considered this
disk, coupled with the Kerr ring, as a model of the KN source
\cite{Isr}. This source was  consistently matched with KN exterior
but resulted in the very exotic superluminal matter of the disk.
Then  Hamity  showed in \cite{Ham} that the disk has to be
relativistically rotating and consisting of the matter with zero
energy density and negative
 pressure which diverge near the Kerr singular ring.
 Finally, L\'opez  \cite{Lop} determined the form of the source, matching
 continuously the KN exterior with the bubble determined by the
 oblate coordinate $r=r_0=e^2/2m .$ The function $H$ vanishes at this $r_0$
  which allowed him to match exterior continuously with the flat interior.
\fn{For $a=0,$ the L\'opes source turns into the Dirac model of
spherical electron \cite{Dir}.} This model was consistent,
however, like other shell-like models, it needed the Poincar\'e
stress compensating the repulsion of the charges distributed over
the shell. Meanwhile, the necessary tangential stress appears
naturally in the domain wall field models, and we exploit it in
the presented model of the KN  source.

\bigskip
{\bf 3. Field  model.}

\bigskip
We consider for the matter fields the Abelian Higgs model with the
Lagrangian \cite{NO} \be {\cal L}_{NO}= -\frac 14 F_\mn F^\mn +
\frac 12 (\cD_\m \Phi)(\cD^\m \Phi)^* + V(r), \label{LNO}\ee
coupled with gravity, so $ \cD_\m = \nabla_\m +ie A_\m $ are to be
covariant derivatives. To form a bubble source we need a domain
wall potential $V(r)$ interpolating smoothly between the external
vacuum state, $V^{ext} = 0,$ and the internal `false' vacuum with
$V^{int} = 0.$ The external vacuum should yield the longrange
external em field of the KN solution, while the interior of the
bubble should be superconducting, leading to a shortrange em field
to provide regularization. It is opposite to the structure of the
NO model, but corresponds to the structure of Witten's field model
for superconducting cosmic strings \cite{Wit}, and we use the
supersymmetric version of this model suggested by Morris
\cite{Mor} with three chiral fields $\Phi^{(i)} = \{\Phi, Z,
\Sigma \}, i=1,2,3,$ controlled by the Lagrangian \be {\cal
L}_{matter}= -\frac 14 F_\mn F^\mn + \frac 12 \sum_i(\cD^{(i)}_\m
\Phi^{(i)})(\cD^{(i) \m} \Phi^{(i)})^* + V \label{L3} ,\ee where
the auxiliary field $Z$ is real and uncharged. We need only one
gauge field $ A_\m $ associated with the chiral field $\Phi^{(1)}
\equiv \Phi ;$  therefore, $F_\mn = A_{\m,\n} - A_{\n,\m} $ and
the corresponding covariant derivations will be $\cD^{(1)}_\m
=\nabla_\m +ie A_\m , \ \cD^{(2)}_\m =\cD^{(3)}_\m =\nabla_\m .$

The potential $V$ is determined via the superpotential $W$ by the
relation $V(r)=\sum _i |\d_i W|^2 ,$ where
   $ \d_1 = \d_\Phi , \ \d_2 = \d_Z , \ \d_3 = \d_\Sigma .$
   The given by Morris
superpotential is \be W= \lambda Z(\Sigma \bar \Sigma -\eta^2) +
(cZ+ \m) \Phi \bar \Phi ,\ee where $c, \ \m, \ \eta, \ \lambda$
are real constants. This potential produces a domain wall
interpolating between the internal  $V^{(int)}=0$ and external
vacuum states $V^{(ext)}=0 ,$ determined by the condition $\d_i W
=0 .$ We use the thin wall approximation, assuming that the depth
of the wall $\xi,$ is much smaller than its position $r_0 ,$ which
yields

(in) \ for $r<r_0$ :  $ \ Z=- \m/c; \ \Sigma=0; \ |\Phi|=
\eta\sqrt{\lambda/c},$

(out) \  for $r>r_0 + \xi  $ :  $ \ Z=0; \ \Phi=0; \ \Sigma=\eta
.$

{\it In the external region} we have $V^{(ext)}=0 ,$ and the
unique nonzero chiral field $\Sigma$ is constant. There survives
only electromagnetic matter, which results in  the
Einstein-Maxwell equations leading to the usual KN solution.

{\it For the interior}, $r<r_0 ,$ we have $V^{(int)}=0 ,$ and the
unique nonzero chiral field is $\Phi .$ As a result, the
Lagrangian (\ref{L3}) is reduced to (\ref{LNO}), leading to the
usual equations for the coupled with gravity Maxwell-Higgs system
\bea \cD^{(1)}_\n \cD^{(1) \n} \Phi &=& \d_{\Phi^*} V = 2 \Phi [
c^2|\Phi|^2
- \lambda c \eta^2]=0 , \label{PhiIn} \\
\nabla _\n \nabla^\n  A_\m &=& I_\m = \frac 12 e |\Phi|^2
(\chi,_\m + e A_\m). \label{AIn} \eea However, as we see in the
next section, there is a strong reason to choose parameters of the
model which lead to the flat space inside the bubble. As a result,
these equations are simplified, allowing us to obtain exact
solutions.

\bigskip
 {\bf 4. Regularization of the KN metric.}

\bigskip
 The KS metrics which describe a smooth interpolation
from the external metric of the KN solution to a regular  internal
space-time where considered in \cite{BurBag,BEHM} on the base of
the  GG ansatz \cite{GG}. For the external region one uses the
exact KN solution in the KS form  \be g_\mn=\eta _\mn + 2 H k_\m
k_\n , \label{ksH} \ee with $H=\frac {mr - e^2/2}{r^2 + a^2 \cos
^2 \theta}.$ The KN electromagnetic (em) field is given by the
vector potential \be A^\m_{KN} = Re \frac e {r+ia \cos \theta}
k^\m , \label{AKN} \ee where $k^\m(x^\m)$ is the null vector field
tangent to a vortex of the null geodesic lines, the Kerr principal
null congruence (PNC). The field $k^\m(x)$ has the form \cite{DKS}
\be k_\m dx^\m = dr - dt - a \sin ^2 \theta d\phi_K . \label{km}
\ee

For metric inside  the bubble, $r<r_0 ,$ one uses the KS form
(\ref{ksH}) with the function \be H=f(r)/\Sigma, \quad \Sigma=(r^2
+ a^2 \cos ^2\theta) \label{HGG} .\ee It describes the external KN
metric for $f(r)=f_{KN}= mr -e^2/2 ,$ and any smooth function
$f(r)$ interpolating between $f_{KN}(r)$ and  $f_{itn}(r)$
provides a smooth transfer to a metric inside the bubble. If we
set for the interior $f(r)=f_{int}=\alpha r^4 ,$ the Kerr
singularity will be suppressed. The GG form of function $H$
({\ref{HGG}) allows one to describe in a universal manner the de
Sitter and Anti-de Sitter solutions, and  match smoothly  the
external and internal metrics of different types. Besides, it
demonstrates a correspondence between  rotating and non-rotating
Kerr-Schild solutions which simplifies analysis of the rotating
solutions. The Kerr space-time is foliated into
 ellipsoidal layers $r=const. $ which are rigidly rotating with
 angular velocities depending on the Kerr oblate coordinate $r,$
$\Omega(r) = \frac a {a^2 +r^2}.$ All the fields on the layers may
be transformed to the co-rotating coordinate systems by the
corresponding local Lorentz transformation, where they take a
simple diagonal form \cite{BurBag,BEHM,GG,RenGra}) \be T_\mn
=diag(\rho, \ p_{tan}, \ p_{tan}, \ p_{rad}), \label{TGG1} \ee
where \be\rho= -p_{rad}=\frac 1{4\pi}(f'r -f)/\Sigma^2 , \
p_{tan}=\rho -\frac 1{8\pi} f''/\Sigma \label{TGG2} ,\ee
 corresponding to the non-rotating solutions. Because of that
many results obtained for the non-rotating case ( $a=0$ and
$\Sigma=r^2$ in (\ref{HGG})), may easily be translated back to the
rotating case by the replacement $\Sigma \to r^2 + a^2 \cos^2
\theta .$ In particular, it allows one to match smoothly the
external and internal metrics with rotation.

Setting $f_{int}=\alpha r^4 ,$ one obtains from (\ref{TGG1}) and
(\ref{TGG2}) that it corresponds  a regular rotating internal
space-time of the constant curvature $R=-24 \alpha ,$
\cite{BurBag,BEHM}.  The energy density inside the bubble will be
$\rho=\frac {3\alpha} {4\pi},$ which generates de Sitter interior
for $\alpha
>0,$ anti de Sitter interior for $\alpha <0 $ and the flat
interior corresponding to the L\'opez model for $\alpha =0 .$
 The three corresponding  branches for the interior are shown
in fig.1 (taken from \cite{RenGra}). They intersect the plot
$f_{KN}(r)$ at the points $r_0=r_{AdS},
 \ r_0=r_{dS}$ and $r_0= r_{flat}$ correspondingly, showing the point where
the external KN metric matches continuously with the interior.
\begin{figure}[ht]
\centerline{\epsfig{figure=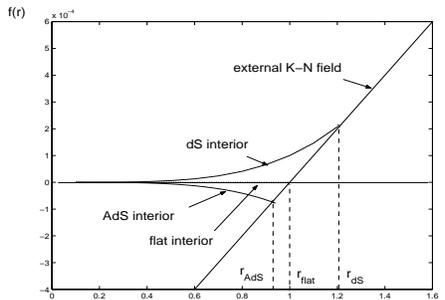,height=4cm,width=6cm}}
\caption{Matching of the metric for regular bubble interior with
metric of external KN field.}
\end{figure}
Selecting a thin layer of phase transition,  $r_0 <r <r_0 + \xi ,$
one can take a smooth interpolating function $f(r)$ and provide
smoothness of the general metric. In the thin domain wall limit
$\xi \to 0 ,$ the point of intersection $f_{int}(r)=f_{KN}(r)$
determines $r_0$ graphically. One sees that $r_0$ is solution of
the equation $\alpha r_0 ^4 =mr_0 -e^2/2 ,$  and since
$\rho=\frac{3\alpha} {4\pi},$ one obtains for ADM mass the
relation \be m = e^2/2r_0 + \rho \frac 43 \pi r_0^3 ,
\label{mass}\ee where the first term on r.h.s. is the mass of the
external em field for the charged sphere, and the second term is
the mass inside  this sphere with the matter density $\rho.$ For
small $\xi$ the contribution of the $\xi$-shell to the mass $m$ is
negligible, and we obtain an important result that {\it the point
of the phase transition $r_0$ is determined by the `balance matter
equation' (\ref{mass})}. This analysis is extended to the rotating
case, in which $r$ has to be considered as the Kerr oblate
ellipsoidal coordinate, and the boundary of bubble takes the form
of an oblate rotating disk, \cite{BurBag,BEHM,RenGra}.

We  restrict further the treatment by the case of a flat interior
 corresponding to $\alpha=\rho=f(r)=0, \ r<r_0 .$

{\it Stringy effect.} In the rotating case the
Arnowitt-Deser-Misner (ADM) mass of the KN source is determined by
$ T^0_0 $ component of the stress-energy tensor in the
asymptotically flat coordinate system.  Contributions to the ADM
mass coming from the interior of the bubble, the intermediate
region, and the external electromagnetic field were calculated in
\cite{BEHM}. For the flat interior, $\alpha=0 ,$ and an infinitely
thin transitional shell we have: $ m_{ADM}^{(int)} =0 ;$ and \be
m_{ADM}^{(shell)}= \frac m2 [ 1- (\frac {r_0} a
 + \frac a {r_0} ) \arctan (\frac a {r_0}) ]; \label{mshell} \ee
 \be m_{ADM}^{(ext)} = \frac m2 [ 1+ (\frac {r_0} a + \frac a
{r_0} ) \arctan (\frac a {r_0}) ] \label{mext} . \ee For  $a>>r_0
,$ the second term in the square brackets of (\ref{mext}) tends to
the expression $ \delta^{string} \approx \frac {a \pi} {2 r_0} $
which may be interpreted as a stringy contribution to the
mass-energy caused by concentration of the electromagnetic field
on the edge of bubble. \fn{In particular, for the parameters of an
electron (in the units $\hbar=c=G=1$) the bubble forms a highly
oblate disk with the radius about the Compton length $a = J/m
=1/2m$ and the thickness equal to the "electromagnetic" radius of
the electron, $r_0=r_e=e^2/2m .$ The degree of oblateness is
$r_0/a = e^2 \sim 137^{-1}.$}
  Stringy structure of the region near the Kerr ring
 was discussed many times, \cite{BurSen,Isr1,BurTwi,BurAli}.
 Assuming that the closed Kerr string has a tension $T $ with
the rising potential $E\equiv m= Ta ,$ one can combine it with the
basic KN relation $J=ma ,$ and obtain the linear Regge trajectory
$J= \frac 1T m^2$ with the slope $1/T .$ However, one sees from
(\ref{mshell}) and (\ref{mext}) that the `stringy' contributions
from the shell and external em field are mutually cancelled, and
the total mass $m_{ADM}^{(total)} = m_{ADM}^{(int)} +
m_{ADM}^{(shell)} + m_{ADM}^{(ext)}$ turns out to be equal to $m
.$ Indeed, this result could be predicted {\it a priori}, since
the total ADM mass is determined {\it only} by the asymptotical
gravitational field, i.e. by the parameter $m$ in function $H ,$
(\ref{HKN}). The treatment of the Tolman mass expression, or the
transitional zone of a finite thickness does not change the
result. The partial stringy contributions to the  mass are very
intense ($\delta m /m > 100$), but mutual cancelling of the
negative and positive  contributions prevents exhibition of the
stringy effect in the considered isolated system. However, the
balance of matter may apparently be destroyed by an external
field, and a considerable effect may be expected in the bound
systems or in consequence of the radiative corrections.

\bigskip
{\bf 5. Regularization of the KN electromagnetic field.}

\bigskip
We assume that for $r<r_0$ the phase transition is completed and
inside the bubble the field $ \Phi(x)$ has the form $ \Phi(x) =
|\Phi(x)|e^{i \chi(x)}$ with a nonzero vev,
$|\Phi(x)|_{r<r_0}=\Phi_0 .$ We have to obtain a regular solution
of equation (\ref{AIn}) for $r<r_0$ in the presumption of the flat
interior, $\alpha=0 ,$ which fixes the boundary of
 bubble at $r_0=r_e=e^2/2m .$ The flat interior
allows us to use the flat d'Alemberian and  $\cD_\m = \d_\m +ie
A_\m $ in (\ref{AIn}) and  yields \be \Box A_\m =I_\m = e |\Phi|^2
(\chi,_\m + e A_\m) \label{Main}.\ee The current has to be
expelled from the bulk of the superconducting interior to the
boundary of the bubble and we should set in the interior $I_\m=0
,$ which yields \be \Box A_\m^{(in)} = 0 = e |\Phi|^2 (\chi,_\m +
e A_\m^{(in)}) \label{MainIn}.\ee

 The external KN  field $A_\m$ is given by (\ref{AKN}) and (\ref{km}).
Its matching with the interior turns out to be nontrivial, since
the Kerr angular coordinate $\phi_K $ (\ref{oblate}) is very
specific, and inside the bubble it turns out to be inconsistent
with the simple angular coordinate of the Higgs field,
$\phi=-i\ln[(x+iy)/\rho],$ where $\rho=(x^2+y^2)^{1/2} .$ Using
the relation between the corresponding differentials \be d\phi_K
=\phi + \frac {adr}{r^2+a^2} ,\ee we transform the KN vector
potential on the boundary and inside the bubble to the form \be
A_\m dx^\m = \frac {-er}{r^2+a^2\cos^2 \theta} [dt + a \sin ^2
\theta d\phi ] + \frac {2e r dr} {(r^2 +a^2)} \label{Aphiin} \ee
which shows that the radial component $A_r$ is a perfect
differential. On the external side of the bubble, $r=r_e +0 =
e^2/2m ,$ the value of the potential is \be A_\m dx^\m|_{r_e +0} =
\frac{-e r_e} {r^2_e +a^2 \cos^2 \theta}[dt + a \sin ^2 \theta
d\phi ] + \frac {2e r_e dr} {(r_e^2 +a^2)} . \ee The lines of the
KN vector-potential in the equatorial plane, $\cos\theta=0 ,$ are
tangent to the Kerr singular ring, \cite{BurSen,BurTwi}, and
approaching the string-like edge of the KN bubble, the potential
takes the form \be A^{(str)}_\m dx^\m = - \frac{2m} {e}[dt + a
d\phi ] + \frac {2e r_e dr} {(r_e^2 +a^2)}. \ee The tangent
component on the edge takes the value $A^{(str)}_\phi= - 2ma/e .$
In agreement with (\ref{Main}) and $I_\m=0 ,$ the component
$A^{(str)}_\phi$ has to be matched with angular periodicity of the
internal Higgs field $\Phi = \Phi_0 \exp(i\chi ),$ which
determines its
 $\phi$-dependence,  $\Phi \sim \exp \{i n \phi \} ,$  $n$-integer.
Therefore, the condition $I_\m=0$ demands mutual compensation of
the gauge and Higgs field inside the bubble and leads to  the
closed Wilson loop on its edge. The integral \be S=\oint
eA^{(str)}_\phi d\phi=-4\pi ma \label{WL} \ee   has to be matched
with incursion of the Higgs phase $2\pi n ,$ which, due to the KN
relation $J=ma ,$ entails $J=n/2,$ where $n$ is integer.
Therefore, the consistent matching selects quantum values of the
KN angular momentum.

 For the time-like component $A_0$ inside the bubble, the condition
 $I_\m=0$ yields $\chi,_0=-eA^{(in)}_0(r),$ which determines
 $\chi,_0$ as a constant corresponding to the frequency of oscillation
 of the Higgs field, $\chi,_0=\omega=-eA^{(str)}_0=2m .$ Therefore, the
vector potential inside the bubble has the components  \be
A_0^{(in)} =-\frac {2m}e; \ A_\phi^{(in)} =-\frac {2ma}e; \
A_r^{(in)} = \frac {2e r dr} {(r^2 +a^2)} . \label{Ain}\ee This
potential satisfies the left-hand side of equation (\ref{MainIn}),
so far as it is gradient of the scalar function $\chi(t,r,\theta)
.$ At the same time, it also satisfies the right-hand side of
(\ref{MainIn}), since it is compensated by the phase of the Higgs
field.

The internal Higgs field takes the form \be \Phi(x)= \Phi_0 \exp
\{i\omega t - i \ln (r^2 +a^2) + 2i n \phi  \} . \label{Hin}\ee
For exclusion of the region of the
 closed loop at the equator, the time and $\phi$ components of
 the obtained vector field have a chock, crossing the boundary of the
 bubble,  which generates circular currents distributed over the
 surface layer.

\bigskip
{\bf 6. Consistency of the chiral model with gravity.}

\bigskip
 The complete stress-energy tensor may be decomposed into pure em
part and contributions of the chiral fields  \bea T^{(tot)}_\mn
= T^{(em)}_\mn + \label{T3} \\
\nonumber   \delta_{i\bar j}(\cD^{(i)}_\m \Phi^i)\overline
{(\cD^{(j)}_\n \Phi^j)} &-& \frac 12 g_\mn[\delta_{i\bar
j}(\cD^{(i)}_\lambda \Phi^i)\overline {(\cD^{(j)\lambda} \Phi^j)}
+V] ,\eea
 In the exterior we have $ V^{(ext)}=0 ,$ and the unique
nonzero chiral field $\Sigma$ tends to constant. Therefore, all
the derivatives $\cD^{(i)}_\m \Phi^{(i)}$ will vanish, and
$T^{(tot)}_\mn = T^{(em)}_\mn ,$ which leads to consistency with
the Einstein-Maxwell equations and to the external KN solution.
For the interior, $r<r_0,$ we have the flat space, $\alpha=0 ,$
and $V^{(int)}=0 ,$ which yields the unique nonconstant Higgs
field $ \Phi(x) = |\Phi(x)|e^{i \chi(x)} .$ The Lagrangian
(\ref{L3}) reduces to (\ref{LNO}), leading to the equations
(\ref{AIn}). For the case of flat interior these equations were
solved explicitly in sec.5. As a result,  the stress-energy tensor
(\ref{T3}) turns out to be consistent with the flat interior, and
with the external KN metric. At the same time, its consistency in
the zone of the phase transition is not established at this stage,
and we may consider the obtained solution only as a thin-wall
approximation.

\bigskip
{\bf 7. Beyond the thin wall approximation.}

\bigskip
 The structure of the zone of the phase transition is very nontrivial and
 deserves especial study. In this section, we give only a preliminary
 treatment of the encountered problems. A thin
but finite zone of the phase transition may be modelled by a
planar vacuum domain wall, the stress-energy tensor of which has
the following typical structure \cite{IpsSik,CvGrifSol} \be
T^{DW}_\mn = \rho \ {\rm diag} ( 1 , \ - 1 , \ - 1 , \ 0),
\label{TplanVacDW} \ee indicating that the wall tension is equal
to energy density. The stress-energy tensor of the considered
chiral field model also has this form in the case of the planar
vacuum domain wall without the em field. However, subtracting from
(\ref{TGG1}) and (\ref{TGG2}) the contributions from the em field,
we obtain the form \be T_\mn = |p_{tan}| \ {\rm diag} ( 0 , - 1 ,
- 1 , 0) \label{TplanDW} ,\ee which differs from the typical
`vacuum' domain wall, indicating the presence of an unaccounted
structure.
 One of
the neglected phenomena is related to the penetration of the em
field in the superconducting region, the zone of `penetration
length' \cite{NO}, where the gauge field acquires a mass via the
Higgs effect. In our case this zone forms an extra layer
$U_{\delta}=[r_0 -\delta < r \le r_0 ]$ on the boundary of the
disk-like bubble. The massive gauge field concentrates on the edge
of bubble in the equatorial plane, and forms  a circular stringy
region positioned very close to the former Kerr singular ring (the
Cartesian distance from the Kerr ring is about $a e^4 $). On the
external side of the bubble, $r=r_0 +0$, this massive field
transfers into the gauge field of the above Wilson loop, while on
the internal side of the bubble it creates circular currents,
which are determined by the deviation of the real em field from
the  background solutions (\ref{Ain}), (\ref{Hin}) obtained in
section 5.
 Substituting $ A_\m^{(\delta)} = A_\m - A_\m^{(in)} $ into (\ref{Main})
 one obtains the equation
 \be \Box A_\m^{(\delta)} =I_\m = e^2 |\Phi|^2  A_\m^{(\delta)} \label{skin}\ee
 which describes  this effect.

The circular stringy region rotates together with the bubble with
angular velocity $\Omega = ca /(a^2 +r_0^2).$ So, its linear
velocity on the edge of bubble is $v_0=c /\sqrt {1+ (r_0/a)^2 }
\sim c .$ In particular, for the parameters of an electron $ r_0/a
= 137^{-1} ,$ and the linear velocity of the edge $v_0 \approx
0.9999 c  $ is almost lightlike. Therefore, we again arrive at the
stringy structure which appears in the transition region. This
structure is very nontrivial and important, and we intend to
considered it in a separate paper.

\bigskip
{\bf 8. Conclusion.}

\bigskip
In the thin-wall approximation, we obtained the regular solution
of the  KN geometry which may be considered as a charged, spinning
and gravitating soliton.  The Higgs field inside its regular core
is coherently oscillating, which reproduces the typical property
of the oscillon models: spinning Q-balls \cite{VolkWohn,Grah} and
bosonic star models \cite{BosStar}.

 One of the new important peculiarities of this solution is
the appearance of the closed Wilson loop at the boundary of the
bubble source which leads to quantization of the total angular
momentum. The closed vortexes were also obtained  in some other
soliton models \cite{Shnir,VolkWohn}, but in our case the Wilson
loop has a specific origin, being determined by the vortex
structure of the Kerr singular ring.

The considered model is Abelian and may be related to the light
spinning particles.  Although, it may apparently be generalized to
the non-Abelian case too. For the parameters of an electron, the
considered regular solution predicts a disk-like core formed by
the Higgs field, which is bounded by a closed circular current of
the Compton radius. In principle, it agrees with the size of a
dressed electron predicted by QED, but suggests some extra
features of the spinning solutions consistent with general
relativity. First is that the core is coherently oscillating and
therefore, it should be considered as an integral element of the
spinning particle.  Second is the disk-like form of the core,
which should apparently have experimentally observable
consequences.\fn{It could be experiments related to very soft
scattering of the polarized electrons. Note that Compton discussed
experimental confirmation of their disklike shape in \cite{Comp}.}

 Finally, we should note that the mysterious problem of twosheetedness
of the KN spacetime is turned into its advantage, since the KN
source forms a holographic structure \cite{BurPreQ} which is
necessary for quantum treatments  \cite{Gib,SHW}.
 The inner flat vacuum state may be extended analytically to the
 negative KN sheet
 which takes the role of a flat in-vacuum space, separated from the physical
 out-sheet by the holographically dual boundary of the bubble, forming the
 bulk/boundary correspondence.

\bigskip
 {\bf Acknowledgements.}

\bigskip
  The author thanks T.M. Nieuwenhuizen
for hospitality in ITP of Amsterdam University and for discussions
which stimulated reconsideration of this problem, and also thanks
D. Stevens for conversations and useful reference.

\end{document}